\documentclass[aps,twocolumn,showpacs]{revtex4}
\usepackage{epsfig}
\usepackage{amsbsy}
\usepackage{amsmath}
\usepackage{amsfonts}
\usepackage{graphicx}

\begin{document}

\newcommand{\bqa}{\begin{eqnarray}}
\newcommand{\eqa}{\end{eqnarray}}
\newcommand{\nn}{\nonumber}
\newcommand{\nl}[1]{\nn \\ && {#1}\,}
\newcommand{\erf}[1]{Eq.\ (\ref{#1})}
\newcommand{\dg}{^\dagger}
\newcommand{\ip}[1]{\left\langle{#1}\right\rangle}
\newcommand{\bra}[1]{\langle{#1}|}
\newcommand{\ket}[1]{|{#1}\rangle}
\newcommand{\braket}[2]{\langle{#1}|{#2}\rangle}
\newcommand{\st}[1]{\langle {#1} \rangle}
\newcommand{\etal}{{\it et al.}}
\newcommand{\half}{{\small \frac 12}}

\title{Near-optimal two-mode spin squeezing via feedback}
\author{Dominic W.\ Berry and Barry C.\ Sanders}
\affiliation{Department of Physics and
	Centre for Advanced Computing --- Algorithms and Cryptography,	\\
	Macquarie University, Sydney, New South Wales 2109, Australia}
\date{\today}

\begin{abstract}
We propose a feedback scheme for the production of two-mode spin squeezing. We
determine a general expression for the optimal feedback, which is also
applicable to the case of single-mode spin squeezing. The two-mode spin squeezed
states obtained via this feedback are optimal for $j=1/2$ and are very close to
optimal for $j>1/2$. In addition, the master equation suggests a Hamiltonian
that would produce two-mode spin squeezing without feedback, and is analogous to
the two-axis countertwisting Hamiltonian in the single-mode case.
\end{abstract}
\maketitle

\section{Introduction}
\label{sec:introduction}
Two-mode spin squeezed (TMSS) states are important to consider for a number of
reasons. First, they may be used to demonstrate entanglement experimentally
using only spin measurements \cite{Julsgaard}. Second, as they are entangled,
they may be used as a resource for quantum information, for example, quantum
teleportation \cite{Kuzmich,Berry}, quantum computing, superdense coding, etc.\
\cite{NC}. In addition, TMSS states are analogous to two-mode squeezed states
for light \cite{Caves85}, which have proven to be useful for quantum information
purposes \cite{Furusawa}.

There are several approaches to produce TMSS states. Recently they have been
produced experimentally for samples of atoms by quantum nondemolition (QND)
measurements of the spin components \cite{Julsgaard}. In Ref.\ \cite{Kuzmich}
it was proposed to produce TMSS states of light by combining the two
Einstein-Podolsky-Rosen (EPR) output modes of an optical parametric oscillator
with coherent light at polarizing beam splitters.

It is also possible to adapt proposals for generating single-mode spin squeezing
to the two-mode case. For example, it has been proposed that single-mode spin
squeezed states can be produced via an adiabatic approach \cite{Sor01}. In this
approach, one starts with a Hamiltonian for which the ground state is easily
achieved, and slowly varies the Hamiltonian towards one that has as its
ground state the optimal spin squeezed state. In principle, this procedure can
be adapted to obtain optimal TMSS states. A drawback to this method is that, in
practice, the appropriate Hamiltonian may be difficult to realize.

Another technique for generating TMSS states that may be generalized from the
single-mode case is feedback. Feedback has proven to be a powerful technique for
state preparation \cite{Wise94,Laura} and measurement \cite{Wise98,Berry01}. It
has been shown that it is possible to produce near-optimal spin squeezing in the
single-mode case by continual measurement of the spin operator $J_z$, and using
feedback to bring $\st{J_z}$ close to zero \cite{Laura}. Here we adapt this
method to the QND measurement approach discussed in Ref.\ \cite{Julsgaard}.

The method for production of TMSS states in Ref.\ \cite{Julsgaard} suffers the
drawback that, although the variances in the sums of the spin components 
are small, their means are not close to zero. As is shown in Ref.\ \cite{Berry},
the optimal TMSS states have zero means. We would therefore wish to obtain
states with zero means in order for them to be as close as possible to
the optimal states. This is analogous to the problem with the production of
single-mode spin squeezing as considered by Thomsen, Mancini, and Wiseman (TMW)
\cite{Laura}.

\section{two-mode spin squeezing}
\label{sec:two}
In Ref.\ \cite{Julsgaard}, TMSS states (conditioned on the measurement record)
are obtained via QND measurements of the spin of the atomic samples. In order to
perform this QND measurement, an off-resonant pulse is transmitted through two
cesium gas samples. The interaction Hamiltonian may be expressed as
\cite{Kuzmich,KMJYEB}
\begin{equation}
\label{Hint}
H_{\rm int} =  \hbar a \int S_z(z,t) J_z(z,t) dz,
\end{equation}
where $a$ is a coupling constant, $S_z(z,t)$ is the $z$ component of the
instantaneous Stokes vector for the light, and $J_z(z,t)$ is the $z$ component
of the continuous spin operator. These variables are explained further in
Appendix \ref{sec:apa}.

In the Schr\"odinger picture $J_z(z,t)$ is independent of $t$, and the total
Stokes vector ${\bf S}(z)$ is independent of $z$. The total time evolution
operator is, therefore,
\begin{align}
U &= \exp\left( -\frac i\hbar \int H_{\rm int} dt \right) \nn \\
&= \exp\left( -ia \int S_z(z,t) J_z(z) dz \, dt \right) \nn \\
&= \exp\left( -ia \int S_z J_z(z) dz \right) \nn \\
&= \exp(-ia S_z J_z).
\end{align}
In Ref.\ \cite{Julsgaard} the light is initially linearly polarized, so the light
is in the maximally weighted $S_x$ eigenstate. In addition, the total spin $s$
is large, so we may apply a SU(2)$\to$HW(2) contraction. That is, we make the
substitutions \cite{Holstein,Berry}
\begin{align}
S_x & \to s-b\dg b, \\
S_y & \to \sqrt{\frac s2}(b+b\dg), \\
S_z & \to \frac 1i \sqrt{\frac s2}(b-b\dg),
\end{align}
where $b\dg b$ is the number operator for photons subtracted from the $x$ linear
polarization mode (and added to the $y$ linear polarization mode). We will
assume that the probe beam is in a coherent spin state oriented in the
$x$ direction (i.e., $x$ linearly polarized), and take the limit $s\to\infty$ so
that this contraction is exact.

Under this contraction, $\left[ S_y, S_z \right]\to is$. This means that, under
the Heisenberg picture, the operator $S_y$ transforms as
\begin{align}
S_y^{\rm out} = S_y^{\rm in} + \left[ ia S_z J_z,S_y \right] 
\approx S_y^{\rm in} + \frac{an}2 J_z,
\end{align}
where $n=2s$ is the total number of photons in the pulse. Now we consider two
samples, where we will use the notation $J_k^{(1)}$ and $J_k^{(2)}$, where
$k\in\{x,y,z\}$, for the spin operators for samples 1 and 2, respectively, and
$J_k^{(\pm)}=J_k^{(1)}\pm J_k^{(2)}$. We may replace $J_z$ with $J_z^{(+)}$, the
total $z$ component of spin for the two samples, giving
\begin{equation}
\label{Sdif}
S_y^{\rm out} \approx S_y^{\rm in} + \frac{an}2 J_z^{(+)}.
\end{equation}
Thus a measurement of $S_y^{\rm out}$ gives a QND measurement of $J_z^{(+)}$. In
considering detection, it is more convenient to consider the Stokes vector for
the light integrated over time interval $\delta t$. The transformation of this
component is
\begin{equation}
\label{Sdel}
\delta S_y^{\rm out} \approx \delta S_y^{\rm in} + \frac{a\delta n}2 J_z^{(+)},
\end{equation}
where $\delta n$ is the total number of photons in time $\delta t$.
In Ref.\ \cite{Julsgaard}, TMSS states are obtained by performing QND
measurements on $J_y^{(+)}$ as well. They achieve this by applying a magnetic
field in the $x$ direction, in order to induce Larmor precession of the $y$ and
$z$ components of the spin. This means that Eq.\ (\ref{Sdel}) becomes
\begin{equation}
\label{instant}
\delta S_y^{\rm out}(t) \approx \delta S_y^{\rm in}(t) + \frac {a \delta n}2
[J_z^{(+)}\cos(\Omega t)+J_y^{(+)}\sin(\Omega t)],
\end{equation}
where $\Omega$ is the frequency of the precession. In this case measurement of
$\delta S_y^{\rm out}$ alternately gives QND measurements of $J_z^{(+)}$ and
$J_y^{(+)}$, thus giving reduced uncertainty in both $J_z^{(+)}$ and
$J_y^{(+)}$.

For the TMSS states considered in Ref.\ \cite{Julsgaard} there is
the requirement that $\st{J_x^{(1)}}=-\st{J_x^{(2)}}\gg 1$. Here we will instead
take the convention that both $\st{J_x^{(1)}}$ and $\st{J_x^{(2)}}$ are large
and close to the total spin $j$, and replace $J_y^{(+)}$ with $J_y^{(-)}$. This
is equivalent to a trivial rotation of coordinates for one of the modes
and does not require a physical alteration to the experiment. If we now define
\begin{equation}
J_z^{(\Omega)}(t) \equiv J_z^{(+)}\cos(\Omega t)+J_y^{(-)}\sin(\Omega t),
\end{equation}
then Eq.\ (\ref{instant}) becomes
\begin{equation}
\delta S_y^{\rm out}(t) \approx \delta S_y^{\rm in}(t) + \frac{a\delta n}2
J_z^{(\Omega)}(t).
\end{equation}

In order for this to be accurate, we require $\delta n$ to be large, so that we
may perform the SU(2)$\to$HW(2) contraction. Under this contraction,
$\delta S_y=\sqrt{\delta s/2}X$, where $X=b+b\dg$. Therefore the transformation
in the quadrature $X$ is
\begin{equation}
\label{quadrature}
X^{\rm out}(t) = X^{\rm in}(t)+a\sqrt{\delta n} J_z^{(\Omega)}(t).
\end{equation}
As the initial state is equivalent to the vacuum state [under the
SU(2)$\to$HW(2) contraction], it has a variance of 1 in its measured value. The
same is true for $X^{\rm out}$, as the extra term only changes the mean.

Measurements are made on the output beam by splitting the beam into $+45^\circ$
and $-45^\circ$ linearly polarized components at a polarizing beam splitter and
measuring the intensities at the two outputs. We then define the photocurrent as
\begin{equation}
\label{photocurrent}
I_c(t) = \lim_{\delta t\to 0} \frac{\delta N_+ -\delta N_-}{\sqrt \nu \delta t},
\end{equation}
where $\delta N_+$ and $\delta N_-$ are the photon counts at the $+45^\circ$ and
$-45^\circ$ linearly polarized outputs and $\nu$ is the photon flux
$\delta n/\delta t$. We may interpret $I_c(t)$ as the measured value of $X$
divided by $\sqrt{\delta t}$, so the photocurrent is given by
\begin{equation}
I_c(t) = a\sqrt{\nu}\st{J_z^{(\Omega)}}_c + \xi(t),
\end{equation}
where the subscript $c$ indicates the average for the conditioned evolution, and
$\xi(t)$ is a real Gaussian white noise term such that
$\st{\xi(t)\xi(t')}=\delta(t-t')$.

We take the limit of small time intervals $\delta t$, though for finite photon
flux this would mean that the SU(2)$\to$HW(2) contraction is no longer valid, as
$\delta n$ would go to zero. In order for the limit to be rigorously correct, we
need to take the limit of large $\nu$ as well as small $\delta t$. Nevertheless,
this limit should be an accurate approximation for large photon flux. The
conditioned master equation is then
\begin{equation}
\label{conditioned}
d\rho_c = \Gamma {\cal D}[J_z^{(\Omega)}]\rho_c dt + \sqrt{\Gamma}dW(t){\cal H}
[J_z^{(\Omega)}]\rho_c ,
\end{equation}
where $\Gamma=a^2\nu/4$, $dW(t)=\xi(t)dt$ is an infinitesimal Wiener increment,
${\cal D}[r]\rho=r\rho r\dg-(r\dg r\rho+\rho r\dg r)/2$, and ${\cal H}[r]
\rho=r\rho+\rho r\dg-{\rm Tr}[(r+r\dg)\rho]\rho$.

\section{Feedback}

The quadrature (\ref{quadrature}) is measured by detecting the photocurrent
(\ref{photocurrent}), thereby yielding TMSS states. The drawback is that this
will yield nonzero values of $\st{J_z^{(+)}}$ and $\st{J_y^{(-)}}$, whereas
for the optimal TMSS states \cite{Berry}, the expectation values are zero. In
order to obtain states closer to the optimal TMSS states, we apply feedback to
bring these expectation values closer to zero. For example, when
$\cos(\Omega t)=1$, we may apply a Hamiltonian proportional to $J_y^{(+)}$ in
order to bring $\st{J_z^{(+)}}$ towards zero, analogous to
the case considered by TMW. When $\sin(\Omega t)=1$, we wish to apply a
Hamiltonian proportional to $J_z^{(-)}$ in order to bring $\st{J_y^{(-)}}$
towards zero. For arbitrary $t$, we will apply the Hamiltonian
\begin{equation}
H_{\rm fb} = F(t)I_c(t),
\end{equation}
where
\begin{equation}
F(t) = \frac {\lambda(t)}{\sqrt\Gamma} J_y^{(\Omega)}(t),
\end{equation}
and
\begin{equation}
J_y^{(\Omega)}(t) \equiv J_y^{(+)}\cos(\Omega t)-J_z^{(-)}\sin(\Omega t).
\end{equation}
We choose this definition for $J_y^{(\Omega)}$ because $[J_z^{(\Omega)},
J_y^{(\Omega)}]=-iJ_x^{(+)}$ [we will omit the $(t)$ from this point for
brevity]. As $\st{J_x^{(+)}}$ is close to $2j$ for the states we are
considering, this means that rotation about
$J_y^{(\Omega)}$ will alter the expectation value of $J_z^{(\Omega)}$.
This Hamiltonian may be implemented by a radio frequency magnetic field
\cite{Sang}. This generates a Hamiltonian proportional to $J_y$, which becomes
$J_y^{(\Omega)}$ due to the Larmor precession of the atomic spin. 

Using this feedback, we obtain the unconditioned master equation \cite{master}
\begin{equation}
\label{master}
\dot\rho = -i[\tfrac 12 \{c\dg F(t)+F(t)c\},\rho]+{\cal D}[c-iF(t)]
\rho,
\end{equation}
with
\begin{equation}
c=\sqrt\Gamma J_z^{(\Omega)}.
\end{equation}
In the single-mode case TMW pointed out that the term $c\dg F(t)+F(t)c$ is
proportional to $J_z J_y+J_y J_z$, which is the two-axis countertwisting
Hamiltonian \cite{Kit}. In a similar way we can derive a Hamiltonian for the
two-mode case that produces TMSS states and is analogous to the two-axis
countertwisting Hamiltonian. It is straightforward to show that
\begin{align}
c\dg F(t)+F(t)c \nn &= \lambda(t)\{[(J_y^{(1)})^2+(J_y^{(2)})^2-
(J_z^{(1)})^2 \nn \\ &~-(J_z^{(2)})^2]\sin(2\Omega t)+(J_z^{(1)}J_y^{(1)}+
J_y^{(1)}J_z^{(1)} \nn \\ & ~~~+J_z^{(2)}J_y^{(2)}+J_y^{(2)}J_z^{(2)})
\cos(2\Omega t) \nn \\ & ~~~+ 2(J_z^{(1)}J_y^{(2)}+J_y^{(1)}J_z^{(2)})\} .
\end{align}
In the limit of large $\Omega$, the contribution to the evolution from the terms
containing $\sin(2\Omega t)$ and $\cos(2\Omega t)$ is negligible. Omitting these
terms yields
\begin{equation}
c\dg F(t)+F(t)c \approx 2\lambda(t)(J_z^{(1)}J_y^{(2)}+J_y^{(1)}J_z^{(2)}),
\end{equation}
which indicates that it is possible to produce two-mode spin squeezing using the
Hamiltonian
\begin{equation}
\label{fouraxis}
H\propto J_z^{(1)}J_y^{(2)}+J_y^{(1)}J_z^{(2)}.
\end{equation}

\subsection{Simple feedback}
Now we will consider the appropriate feedback strength, $\lambda(t)$. It is
easily shown that the variation in $\st{J_z^{(\Omega)}}_c$ due to the
conditioned evolution (\ref{conditioned}) is
\begin{equation}
{\rm Tr}(J_z^{(\Omega)}d\rho_c) = 2\sqrt{\Gamma} dW [
\st{(J_z^{(\Omega)})^2}_c-\st{J_z^{(\Omega)}}_c^2].
\end{equation}
The variation in $\st{J_z^{(\Omega)}}_c$ due to the feedback is
\begin{align}
i\st{[H_{\rm fb},J_z^{(\Omega)}]}_c &=i\frac{\lambda}{\sqrt{\Gamma}}I_c
\st{[J_y^{(\Omega)},J_z^{(\Omega)}]}_c \nn \\
& = -\frac{\lambda}{\sqrt{\Gamma}}I_c \st{J_x^{(+)}}_c.
\end{align}
If $\st{J_z^{(\Omega)}}_c=0$, then the appropriate feedback to keep this equal
to zero is
\begin{equation}
\label{cofeed}
\lambda(t)=\frac{2\Gamma \st{(J_z^{(\Omega)})^2}_c}{\st{J_x^{(+)}}_c}.
\end{equation}
If the unconditioned state is close to being pure, then we may also use this
expression with the unconditioned averages without significant loss of accuracy:
\begin{equation}
\label{feed}
\lambda(t)=\frac{2\Gamma \st{(J_z^{(\Omega)})^2}}{\st{J_x^{(+)}}}.
\end{equation}

Note that this result is comparable to that found for the case of feedback for
single-mode spin squeezing by TMW:
\begin{equation}
\label{singlefeed}
\lambda(t)=\frac{2\Gamma \st{J_z^2}}{\st{J_x}}.
\end{equation}
This is due to the fact that the commutation relations for $J_x$, $J_y$, and
$J_z$ are identical to those for $J_x^{(+)}$, $J_y^{(\Omega)}$, and
$J_z^{(\Omega)}$.

\subsection{Analytic feedback}
\label{analfeed}
Next we will consider the possibilities for using an analytic expression for the
feedback strength $\lambda(t)$. In Ref.\ \cite{Laura} the approximate relations
\begin{align}
\st{J_z^2} & \approx (4\Gamma t+2/j)^{-1}, \\
\st{J_x} & \approx j\exp(-\Gamma t/2),
\end{align}
are used to derive the analytic feedback scheme
\begin{equation}
\label{singanalyt}
\lambda(t)=\frac{\Gamma \exp(\Gamma t/2)}{1+2j\Gamma t}.
\end{equation}
In the two-mode case, it is possible to obtain the corresponding
relation for $\st{J_x^{(+)}}$, but not for $\st{(J_z^{(\Omega)})^2}$.

In these derivations it is convenient to define the scaled variables
$v=\Gamma t$ and $\Lambda(t)=\lambda(t)/\Gamma$. In terms of these variables the
master equation (\ref{master}) becomes
\begin{align}
\label{scmaster}
\dot\rho &= -\frac{i\Lambda(t)}2[ (J_z^{(\Omega)}
J_y^{(\Omega)}+J_y^{(\Omega)}J_z^{(\Omega)}),\rho]+{\cal D}[J_z^{(\Omega)}
\nn \\ &~~~ -i\Lambda(t) J_y^{(\Omega)} ] \rho,
\end{align}
where the dot indicates a derivative with respect to $v$. Using these variables,
the evolution is independent of $\Gamma$. The value of $\Gamma$ does not
qualitatively affect the evolution and merely provides a scaling to the time and
to $\lambda(t)$.

It is straightforward to show that
\begin{align}
\frac{d}{dv}\st{J_x^{(+)}} & = {\rm Tr}(J_x^{(+)}\dot\rho)
\nn \\ & = -\tfrac 12 \st{J_x^{(+)}} + 2\Lambda
\st{(J_z^{(\Omega)})^2} - \tfrac 12\Lambda^2 \st{J_x^{(+)}} \nn \\
& = - \tfrac 12 \st{J_x^{(+)}} + \frac{2\st{(J_z^{(\Omega)})^2}^2}
{\st{J_x^{(+)}}}.
\end{align}
Initially this derivative is equal to zero. However, the value of
$\st{(J_z^{(\Omega)})^2}$ falls rapidly and that of $\st{J_x^{(+)}}$ only falls
slowly, so the second term quickly becomes negligible. If we neglect the second
term, the evolution of $\st{J_x^{(+)}}$ is
\begin{equation}
\st{J_x^{(+)}} \approx 2j\exp(-v/2),
\end{equation}
which is equivalent to that in the single-mode case. This technique fails to
find the corresponding expression for $\st{(J_z^{(\Omega)})^2}$, but we have
found numerically that a good approximation is given by
\begin{equation}
\st{(J_z^{(\Omega)})^2} \approx \frac{e^{-v/4}}{2v+1/j}.
\end{equation}
Using these results in the expression for the feedback (\ref{feed}), we obtain
\begin{equation}
\label{analyt}
\lambda(t)=\frac{\Gamma \exp(\Gamma t/4)}{1+2j\Gamma t}.
\end{equation}

\subsection{Optimal feedback}
\label{sec:optimal}
The third alternative that we will consider is optimal feedback. In order to
derive this, it is convenient to define the variables
\begin{equation}
\zeta =\frac {\st{(J_z^{(+)})^2+(J_y^{(-)})^2}}{2j}, ~~~~~~
\chi =\frac {\st{J_x^{(+)}}}{2j}.
\end{equation}
The initial state used is that with individual coherent spin states in each arm,
so both $\zeta$ and $\chi$ are equal to 1. As time progresses both $\chi$ and
$\zeta$ decrease. In order for the feedback to be optimal, the value of $\zeta$
should be as small as possible for each value of $\chi$. Therefore, if $\chi$
decreases by an amount $\delta\chi$, then the amount by which $\zeta$ decreases
should be maximized. This means that the feedback that is optimum will be that
which produces the maximum slope. Therefore, in order to determine the optimal
feedback, we wish to solve
\begin{equation}
\frac{d}{d\Lambda}\frac{\partial\zeta}{\partial\chi} = 0.
\end{equation}
This can be solved using
\begin{equation}
\label{solve}
\frac{d}{d\Lambda}\left[\frac{\frac{d}{dv}\st{(J_z^{(+)})^2+(J_y^{(-)})^2}}
{\frac{d}{dv}\st{J_x^{(+)}}}\right] = 0.
\end{equation}

Using the master equation (\ref{master}), it is straightforward to show that
\begin{equation}
{\rm Tr}\big[(J_x^{(+)})\dot\rho\big] = -\tfrac 12 [1+\Lambda^2]\st{J_x^{(+)}}+
2\Lambda\st{(J_z^{(\Omega)})^2}.
\end{equation}
Determining the corresponding expression for the numerator in Eq.\ (\ref{solve})
is more difficult, and if it is performed directly it leads to an extremely
complicated expression. However, there is a more convenient way of determining
this expression. Note that
\begin{align}
\st{(J_z^{(\Omega)})^2}\! &= \!\langle (J_z^{(+)})^2\!\cos^2(\Omega t)
\!+\!(J_y^{(-)})^2\!\sin^2(\Omega t)\!+\!(J_y^{(-)}\!J_z^{(+)} \nn \\ &~~~
+J_z^{(+)}J_y^{(-)})\cos(\Omega t)\sin(\Omega t)\rangle .
\end{align}
For large $\Omega$ we may average over $t$, which gives
\begin{equation}
\st{(J_z^{(\Omega)})^2} \approx \tfrac 12 \st{(J_z^{(+)})^2+(J_y^{(-)})^2}.
\end{equation}
Numerically this approximation was found to be very accurate. Using this result
we find that
\begin{align}
& \tfrac 12 {\rm Tr}[((J_z^{(+)})^2 +(J_y^{(-)})^2)\dot\rho]
 \approx {\rm Tr}[(J_z^{(\Omega)})^2\dot\rho] \nn \\
& =\Lambda^2 \langle(J_x^{(+)})^2-(J_z^{(\Omega)})^2\rangle
 - \Lambda \langle 4 J_z^{(\Omega)}J_x^{(+)}
J_z^{(\Omega)}+J_x^{(+)}\rangle .
\end{align}

Thus we find that Eq.\ (\ref{solve}) becomes
\begin{equation}
\frac{d}{d\Lambda}\left\{\frac{\Lambda^2 d - \Lambda e}
{-f-\Lambda^2 f +\Lambda g}\right\} =0,
\end{equation}
where
\begin{align}
d & = \st{(J_x^{(+)})^2-(J_z^{(\Omega)})^2}, \nn \\
e & = \st{4 J_z^{(\Omega)}J_x^{(+)}J_z^{(\Omega)}+J_x^{(+)}}, \nn \\
f & = \tfrac 12 \st{J_x^{(+)}}, \nn \\
g & = 2 \st{(J_z^{(\Omega)})^2} .
\end{align}
Solving this gives
\begin{align}
\label{optfeed}
\Lambda = \frac{-fd\pm\sqrt{(fd)^2+ef(fe-dg)}}{fe-dg}.
\end{align}
Numerically it is found that the correct solution that maximizes the slope is
that with the positive sign.

Note that this derivation of the optimum feedback does not rely on any
commutation relations other than the usual su(2) commutation relations. It
therefore is also applicable to the case of feedback for single-mode spin
squeezing, with $J_x^{(+)}$, $J_y^{(\Omega)}$, and $J_z^{(\Omega)}$ replaced by
$J_x$, $J_y$, and $J_z$.

\section{Numerical results}
\label{sec:numer}
\subsection{Unconditioned master equation}
The unconditioned master equation (\ref{scmaster}) was solved using a simple,
finite step method with step sizes of $\delta v=1/1000$. It was found that
better results were obtained as $\Omega$ was increased. In order to estimate the
best result in the limit of large $\Omega$, $\Omega$ was assigned the
maximum value possible with this step size, $\pi/2\delta v$. The initial
condition used was that where the individual modes were in independent coherent
spin states oriented along the $x$ axis.

\begin{figure}
\centering
\includegraphics[width=0.45\textwidth]{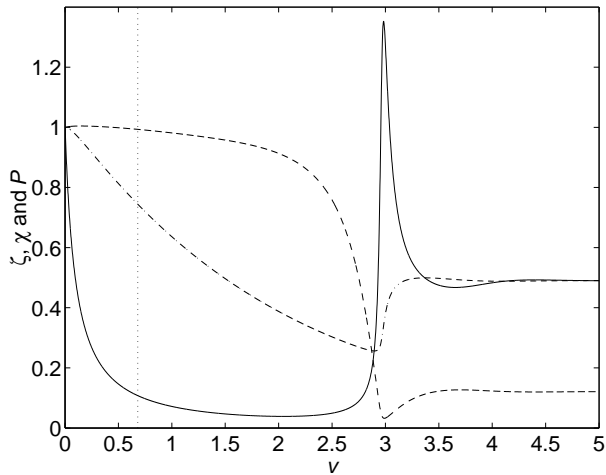}
\caption{The values of $\zeta$ (continuous line), $\chi$ (dash-dotted line), and
the purity (dashed line) as a function of time for simple feedback (\ref{feed})
with a spin of $j=5$. The vertical dotted line indicates the time at which the
state is close to the optimal state considered for teleportation in
Ref.\ \cite{Berry}.}
\label{time}
\end{figure}

At each time step the quantities $\zeta$, $\chi$, and the purity
\begin{equation}
P={\rm Tr}[\rho^2]
\end{equation}
were calculated. The results for the simple feedback scheme of Eq.\ (\ref{feed})
and a spin of $j=5$ are shown in Fig.\ \ref{time}.
Initially the system behaves as we would expect, with the sum of the variances
$\zeta$ decreasing, until a time of approximately $v=3$. Then the variances
dramatically increase, and the purity drops almost to zero. After this, however,
the system stabilizes. Note also that the value of $\chi$ drops regularly as
time increases until the change at $v=3$.

\begin{figure}
\centering
\includegraphics[width=0.45\textwidth]{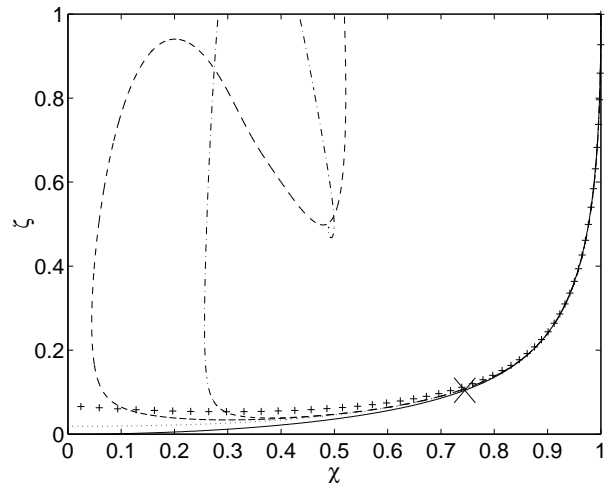}
\caption{The values of $\zeta$ plotted against $\chi$ for spin $j=5$. The
relation for optimal states is shown as the continuous line, the results for the
simple feedback of Eq.\ (\ref{feed}) as the dash-dotted line, the analytic
feedback of Eq.\ (\ref{analyt}) as the dashed line, optimal feedback as the
dotted line, and the countertwisting Hamiltonian (\ref{fouraxis}) as the pluses.
The cross shows the position of the optimal state considered for teleportation
in Ref.\ \cite{Berry}.}
\label{xichi}
\end{figure}

Recall that the system must be entangled if $\st{(J_z^{(+)})^2+(J_y^{(-)})^2}
<\st{J_x^{(+)}}$ \cite{Berry}, which is equivalent to $\zeta<\chi$. Therefore a
state with $\zeta<\chi$ can be described as TMSS. Evidently the feedback
produces quite strong spin squeezing. In order to determine how close to optimal
the states produced by this feedback are, the value of $\zeta$ was plotted
against $\chi$ and compared with the plot for optimal TMSS states in
Fig.\ \ref{xichi}. As time progresses the state travels from the upper right
corner towards the lower left corner of the figure. 

The state starts out very close to optimum and does not deviate significantly
from optimum until $\zeta<0.4$. Eventually the value of $\zeta$ increases
dramatically, and the system spirals towards its equilibrium state at
$(0.5,0.5)$. The results for $j=5$ are typical, and similar results are obtained
for other spins $j>1/2$.

For the purposes of teleportation, the important issue is how close it is
possible to get to the optimal TMSS states considered for teleportation in
Ref.\ \cite{Berry}. This state is indicated by the cross in Fig.\ \ref{xichi}.
As can be seen, the states obtained by feedback closely approach this optimal
state. Similar results are obtained for higher spin, but for smaller spin the
states obtained by feedback are further from the TMSS states considered for
teleportation.

The plots of $\zeta$ versus $\chi$ for analytic feedback and optimal feedback are
also shown in Fig.\ \ref{xichi}. The results for analytic feedback are very
similar to those for simple feedback, with $\zeta$ dramatically increasing at
later times. In contrast, under optimal feedback the value of $\zeta$
monotonically decreases to an asymptotic value for $\chi=0$. For small times the
three feedback schemes produce very similar results, and the states obtained
using optimal feedback do not pass significantly closer to the optimal state for
teleportation.

The last alternative that we consider in this section is the Hamiltonian
(\ref{fouraxis}), which is analogous to the two-axis countertwisting
Hamiltonian in the single-mode case. The results for this Hamiltonian are also
shown in Fig.\ \ref{xichi}. For this case there is significant spin squeezing,
but the squeezing is generally poorer than for feedback. Nevertheless, at later
times the value of $\zeta$ does not rise dramatically (until $\chi$ falls below
zero), as opposed to the results for the simple or analytic feedback.

\subsection{Conditioned master equation}
Next we consider the results for the conditioned master equation. There are two
examples where it is necessary to consider the conditioned master equation. In
the case of simple feedback, at the time that $\zeta$ increases dramatically the
purity of the state has greatly decreased. This means that the feedback based on
the unconditioned averages, Eq.\ (\ref{feed}), is a poor approximation of the
feedback based on conditioned averages, Eq.\ (\ref{cofeed}), and will not
accurately keep the average $\st{J_z^{(\Omega)}}_c$ equal to zero. For this
reason we may obtain better results using the conditioned master equation
and the feedback (\ref{cofeed}).

\begin{figure}
\centering
\includegraphics[width=0.45\textwidth]{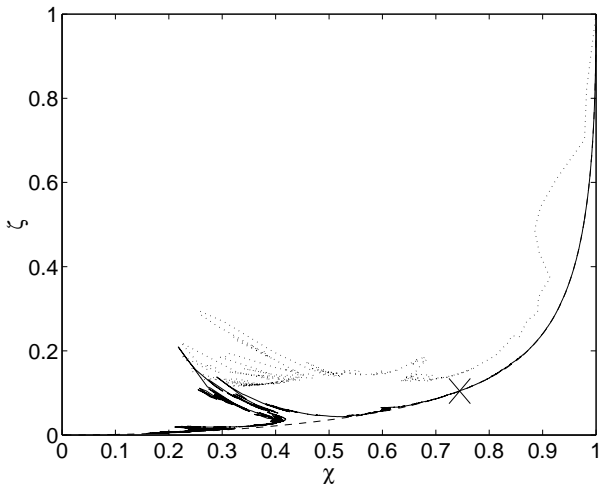}
\caption{The values of $\zeta$ plotted against $\chi$ for spin $j=5$. The
relation for optimal states is shown as the dashed line, the results for
the feedback of Eq.\ (\ref{cofeed}) are shown as the continuous line, and without
feedback as the dotted line. All results are for the conditioned master
equation. The cross shows the position of the optimal state considered for
teleportation in Ref.\ \cite{Berry}.}
\label{figcond}
\end{figure}

The second case that we consider here is that without any feedback. In this case
there is no reduction in the variances for the unconditioned equation. If we
consider the conditioned equation, however, there is a reduction in the
variances, though the means $\st{J_z^{(+)}}_c$ and $\st{J_y^{(-)}}_c$ are
nonzero. For this reason we will consider the conditioned master equation for
this case, with $\zeta$ defined by
\begin{equation}
\zeta =\frac {\st{(J_z^{(+)}-\st{J_z^{(+)}}_c)^2
+(J_y^{(-)}-\st{J_y^{(-)}}_c)^2}_c}{2j}.
\end{equation}
To minimize the random differences between the two cases the same random numbers
were used for each. The results for these two cases are plotted in
Fig.\ \ref{figcond}.

The results for the feedback (\ref{cofeed}) are extremely close to the line
for optimal states, with only temporary excursions from it. In particular, the
value of $\zeta$ does not increase dramatically at large times. This indicates
that improved results may be obtained using feedback based on the conditioned
state.

On the other hand, implementing this feedback may be far more difficult.
In the forms of feedback considered for the unconditioned equation, the feedback
was proportional to the measurement record, with a proportionality constant
$\lambda$ that is a function of time only. This feedback may be performed
efficiently by calculating $\lambda(t)$ beforehand and programming it into a
field programmable gate array \cite{FPGA}. In contrast, the value of $\lambda$
given by Eq.\ (\ref{cofeed}) is dependent on the measurement record, and
therefore would need to be calculated during the measurement, introducing a
significant time delay.

The results without feedback quickly diverge from the line for optimal states,
although they do not differ from it greatly. The major difference is at later
times. While, with feedback, the value of $\zeta$ is reduced very close to zero,
without feedback the value of $\zeta$ cannot be reduced as far. This means that,
although a significant reduction in the variances can be achieved via feedback
for strong QND measurements (where $\Gamma t>1$), feedback does not give much
improvement for weak measurements.

\section{Comparison with single-mode case}
\label{sec:compare}
There is clearly a great similarity between the single-mode and two-mode cases.
The operators $J_x^{(+)}$, $J_y^{(\Omega)}$, and $J_z^{(\Omega)}$ obey exactly
the same commutation relations as $J_x$, $J_y$, and $J_z$ in the single-mode
case. This means that, for example, the optimal feedback also applies to the
single-mode case (as was mentioned in Sec.\ \ref{sec:optimal}). As we may
replace the sum of the variances $\st{(J_z^{(+)})^2+(J_y^{(-)})^2}$ with
$2\st{(J_z^{(\Omega)})^2}$, it might appear that this case is identical to the
single-mode case. Nevertheless, there is a subtle difference due to the fact
that the operator $J_z^{(\Omega)}$ is time dependent.

For the case of spin 1/2, there is complete equivalence between the single-mode
case (for spin $j=1$) and the two-mode case. We will take the Hilbert space for
the single-mode case to be that for two spin 1/2 particles, so that it is the
same as for the two-mode case. The optimal TMSS states for $j=1/2$ are also
optimal single-mode squeezed states in terms of the total spin. That is, they
minimize $\st{(J_z^{(+)})^2}$ for a given $\st{J_x^{(+)}}$. The converse is also
true: the states optimized for single-mode spin squeezing are automatically
optimized for two-mode spin squeezing.

\begin{figure}
\centering
\includegraphics[width=0.45\textwidth]{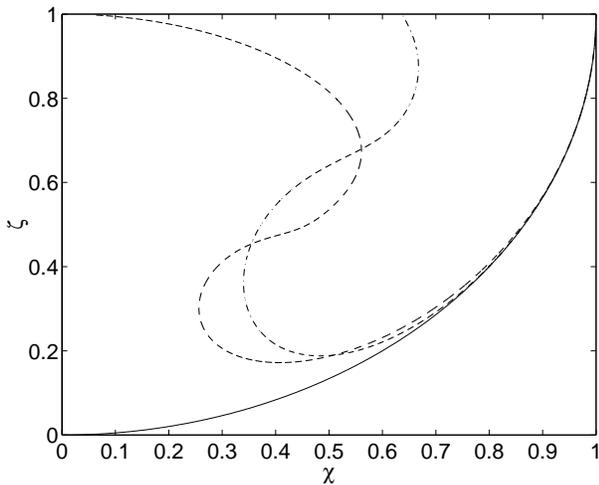}
\caption{The values of $\zeta$ plotted against $\chi$ for spin $j=1/2$ in the
two-mode case and $j=1$ in the single-mode case. The relation for optimal
states, single- or two-mode countertwisting, the feedbacks of Eq.\
(\ref{singlefeed}) in the single-mode case and Eq.\ (\ref{feed}) in the two-mode
case, and the optimal feedback of Sec.\ \ref{sec:optimal}, are shown as the
continuous line. The results for the analytic feedback of Eq.\ (\ref{analyt}) in
the two-mode case, and Eq.\ (\ref{singanalyt}) in the single-mode case are shown
as the dashed line and dash-dotted line, respectively.}
\label{xichi12}
\end{figure}

In addition, both the single- and two-mode countertwisting Hamiltonians produce
optimal spin squeezed states. Not only this, but the feedback as given by
Eq.\ (\ref{singlefeed}) in the single-mode case, or Eq.\ (\ref{feed}) in the
two-mode case, and the optimal feedback described in Sec.\ \ref{sec:optimal},
give optimal spin squeezed states. The only feedback that does not give optimal
states is the analytic feedback for the single- or two-mode case. These results
are depicted in Fig.\ \ref{xichi12}. In this figure the variables $\zeta$ and
$\chi$ are defined for the single-mode case as
\begin{equation}
\zeta = 2\st{J_z^2}/j, ~~~~~~ \chi = \st{J_x}/j.
\end{equation}

To explain these results analytically, first note that, for spin 1/2,
\begin{equation}
J_z^{(+)}J_y^{(+)}+J_y^{(+)}J_z^{(+)}=2(J_z^{(1)}J_y^{(2)}+J_y^{(1)}J_z^{(2)}).
\end{equation}
This means that the countertwisting Hamiltonians for the one- and two-mode cases
are identical, and therefore produce identical states.

In addition we find that $(J_z^{(+)})^2-(J_y^{(-)})^2$ commutes with these
Hamiltonians. This means that for the states produced by countertwisting, the
values of $\st{(J_z^{(+)})^2}$ and $\st{(J_y^{(-)})^2}$ will be identical. This
means that, if the states produced are optimal single-mode spin squeezed states
(i.e., $\st{(J_z^{(+)})^2}$ is minimized), then they must also be optimal TMSS
states.

To show that optimal single-mode spin squeezed states are produced by the
countertwisting Hamiltonian, note that for total spin 1 we have the differential
equations
\begin{align}
\frac{d}{dv} \st{J_x} & = 2\st{J_z^2-J_y^2}, \\
\frac{d}{dv} \st{J_z^2} & = -\st{J_x}, \\
\frac{d}{dv} \st{J_y^2} & = \st{J_x}.
\end{align}
The solution of these equations is
\begin{equation}
\label{optrel}
\st{J_x}^2 = 4(\st{J_z^2}-\st{J_z^2}^2).
\end{equation}
This is the relation for optimal single-mode spin squeezed states as given by
S{\o}rensen and M{\o}lmer \cite{Sor01}. This shows that the states produced by
the one- and two-mode countertwisting Hamiltonians and the one- and two-mode
optimal states are all identical.

For the case of feedback, it is sufficient to show that the feedbacks of Eqs.\ 
(\ref{singlefeed}) and (\ref{feed}) give optimal states, as this will imply that
the optimal feedback gives optimal spin squeezed states. The derivation in this
case is lengthy, and is given in Appendix \ref{derivs}. It is interesting to
note that in this case the optimal feedback is given by Eq.\ (\ref{apeq}), and
differs dramatically from the analytic feedback used for higher spin.

\begin{figure}
\centering
\includegraphics[width=0.45\textwidth]{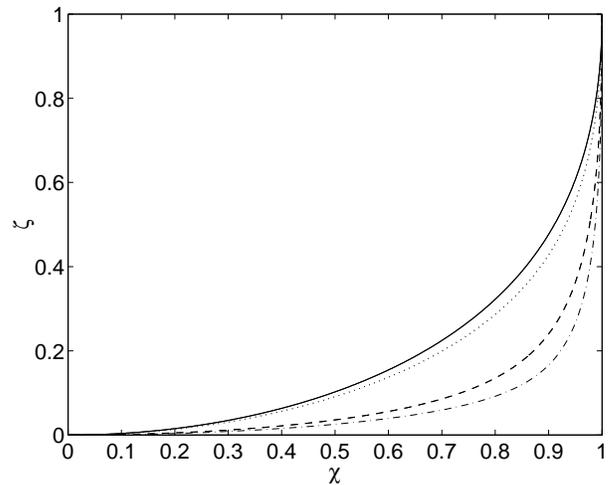}
\caption{The values of $\zeta$ for optimal states plotted against $\chi$ for
total spin 2 for the two-mode case (continuous line) and the single-mode case
(dotted line), and for total spin 10 for the two-mode case (dashed line) and
the single-mode case (dash-dotted line).}
\label{compare}
\end{figure}

The equivalence between the single- and two-mode cases does not hold for any
total spin higher than 1. As shown in Fig.\ \ref{compare}, there are significant
differences between the optimal single- and two-mode spin squeezed states for
total spin above 1. As can be seen, for a total spin of 2 or more the values of
$\zeta$ are significantly less for the single-mode case than for the two-mode
case. This reflects the fact that in the two-mode case we wish to minimize
$\st{(J_y^{(-)})^2}$ as well as $\st{(J_x^{(+)})^2}$.

As there are such strong similarities between the single-mode and two-mode
cases, it is reasonable to apply the squeezing parameter as considered in
Refs.\ \cite{Laura,Sorensen,Wang} to the two-mode case. In the single-mode case
the squeezing parameter was defined by
\begin{equation}
\xi_z^2 = 2j\st{J_z^2}/\st{J_x}^2.
\end{equation}
With the definitions of $\zeta$ and $\chi$ for the single-mode case above,
this can be expressed as $\xi_z^2 = \zeta/\chi^2$. It would seem reasonable to
use the same definition for the two-mode case.

\begin{figure}
\centering
\includegraphics[width=0.45\textwidth]{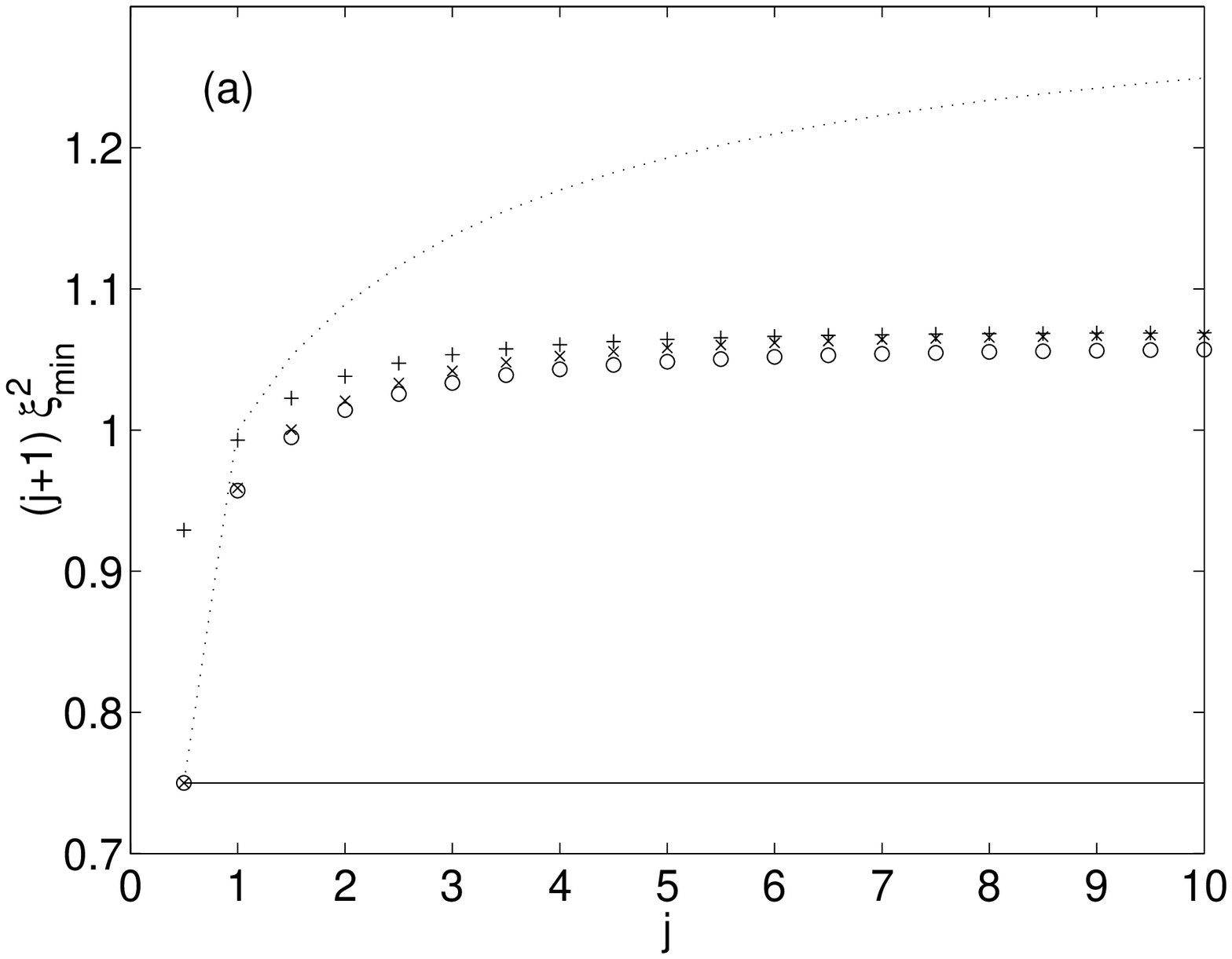}
\includegraphics[width=0.45\textwidth]{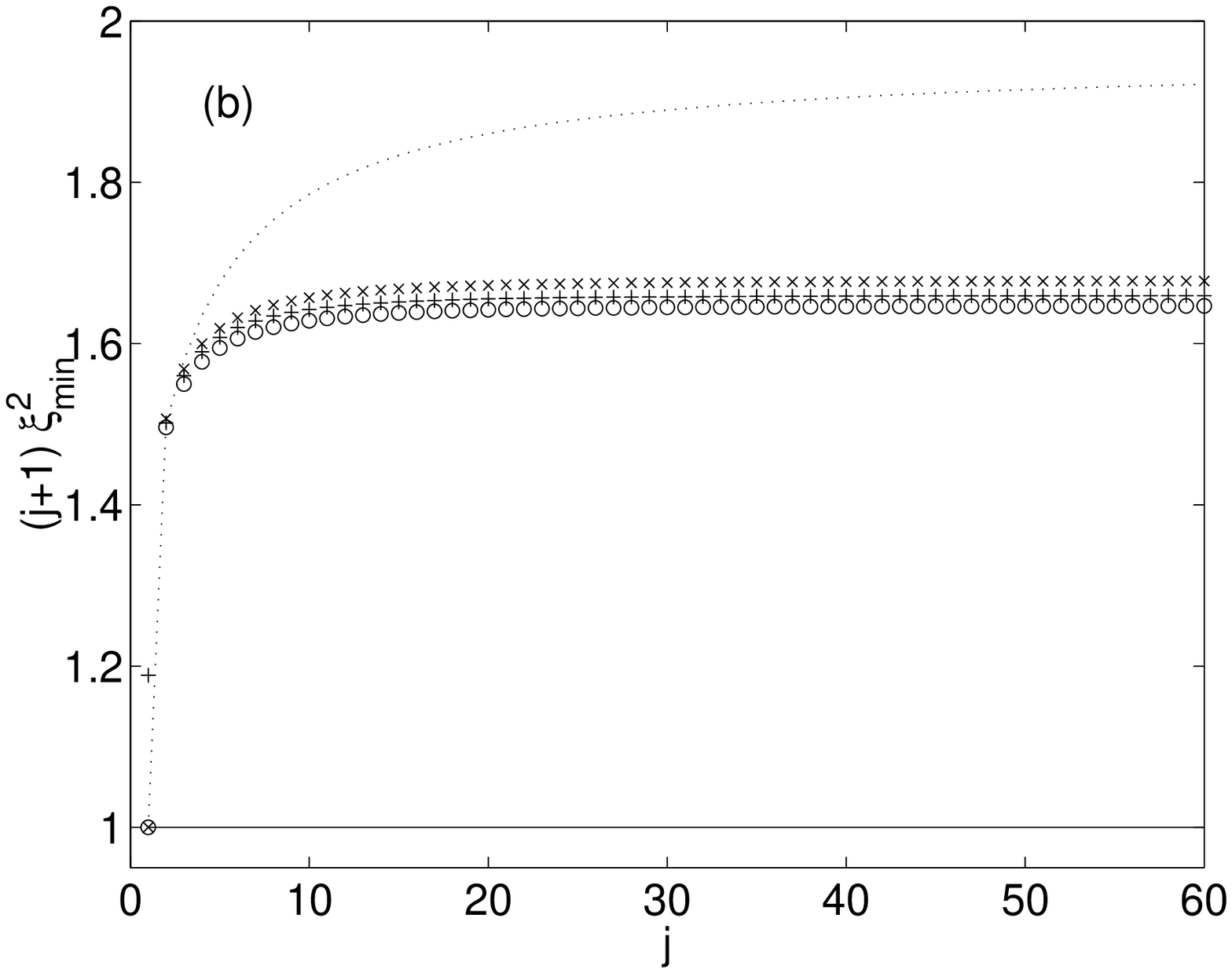}
\caption{The minimum squeezing parameter $\xi_{\rm min}^2$ multiplied by $j+1$
as a function of $j$ in the two-mode case (a) and the single-mode case (b) for
optimal states (continuous line), the feedbacks of Eqs.\ (\ref{feed}) and
(\ref{singlefeed}) for (a) and (b), respectively (crosses), analytic feedback
(pluses), optimal feedback (circles), and countertwisting (dotted line).}
\label{xivar}
\end{figure}

For the two-mode case we find that the minimum value of $\xi_z^2$, namely
$\xi_{\rm min}^2$, varies with spin $j$ as shown in Fig.\ \ref{xivar}(a). In
this figure the values of $\xi_{\rm min}^2$ have been multiplied by $j+1$ in
order to more easily compare the results for different spins. Very similar
results are obtained for the simple feedback of Eq.\ (\ref{feed}) and the
optimal feedback of Eq.\ (\ref{optfeed}). Using the analytic feedback of Eq.\ 
(\ref{analyt}) gives results that are similar to, but slightly above, those for
the other two feedback schemes. All of these feedback schemes give results that
are significantly above the result for the optimal states. Using the
countertwisting Hamiltonian (\ref{fouraxis}) gives results that are higher than
those for feedback.

Similar results are obtained for the single-mode case [see Fig.\ 
\ref{xivar}(b)]. The results for feedback are noticeably above those for
optimal states, and the results for two-axis countertwisting are significantly
above those for feedback. To summarize, the scaling constants for each of the
cases are as given in Table \ref{scales}. In this table we can see the
similarities between the single-mode and two-mode cases: the scaling constants
obtained by feedback are significantly above those for optimal states, and the
scaling constants for two-axis countertwisting (or its two-mode equivalent)
are significantly above those for feedback. In both cases there are only small
differences between the results for the different types of feedback. The only
qualitative difference between the scaling constants for the single- and
two-mode cases are that in the single-mode case the analytic feedback gives
slightly better results than the simple feedback of Eq.\ (\ref{feed}). In
contrast, in the two-mode case the scaling constants are indistinguishable.

\begin{table*}
\caption{The scaling constants for the minimum squeezing parameter
$\xi_{\rm min}^2$ for both single- and two-mode spin squeezing, for optimal
states, simple feedback, analytic feedback, optimal feedback, and two-axis
countertwisting (or its equivalent in the two-mode case). \label{scales}}
\begin{tabular}{lccccc}
	\hline \hline
	  & ~Optimal~ & ~Optimal feedback~ & ~Analytic feedback~ & ~Simple feedback~ & ~Countertwisting~ \\
	\hline
	Single mode & 1   & 1.6468 & 1.6593 & 1.6777 & 1.9562 \\
	Two mode    & 3/4 & 1.0584 & 1.0692 & 1.0692 & 1.292 \\
	\hline \hline
\end{tabular}
\end{table*}

\section{Experimental prospects}
There are some complications in applying this theory to the experiment of
Ref.\ \cite{Julsgaard}. In this experiment, $a\approx 5\times 10^{-13}$, and
$\nu\approx 2\times{10^{16}}{\rm s}^{-1}$. This means that $\Gamma\approx{1.4}
\times{10^{-9}}{\rm s}^{-1}$, which implies that the time required for maximal
squeezing is on the order of $10^9{\rm s}$ (over 20 yr). Therefore a far more
intense beam or a stronger interaction would be required to obtain the two-mode
spin squeezing described here.

It must be emphasized that this long time is required for the measurements, not
the feedback. Feedback may be applied to the states obtained in
Ref.\ \cite{Julsgaard} with a negligible increase in the time required for the
experiment. Nevertheless, for the conditions of this experiment, although the
feedback will bring the means of $J_z^{(+)}$ and $J_y^{(+)}$ towards zero, it
will not significantly reduce the variances. It is only when the QND
measurement can be performed strongly enough (i.e., with a strong enough
interaction over a long enough time) to obtain spin squeezing close to maximum
that an improvement in the variance is obtained by using feedback.

Another problem is spontaneous losses due to absorption of QND probe light. The
loss rate due to this is $N\gamma g^2n/4\Delta^2$, where $N=4j$ is the total
number of atoms and $g$ is the one-photon Rabi frequency. As in the case of
single-mode spin squeezing \cite{Laura}, this loss will be very large over the
time period $1/\Gamma$ for free space. Both this problem and the problem of
the long interaction time may be overcome by performing the experiment in a
cavity in the strong-coupling regime.

Two other common experimental problems are inefficient detectors and time
delays. As for the case of single-mode spin squeezing, inefficient detectors do
not affect the scaling. The system has the potential to be far more sensitive to
time delays than in the single-mode case, however. The problem is that, as the
spin component that is being measured is rotating rapidly, the feedback may be
correcting for measurements of a different spin component.

In order to correct the right spin component, the rotation of the spin component
that is measured should have completed an integral number of rotations during
the time delay. Experimentally, this means that the magnetic field should be
adjusted such that the frequency $\Omega$ is an integral multiple of
$2\pi/\tau$, where $\tau$ is the time delay. Provided that this is done, time
delays should not be more of a problem than in the single-mode case.

\section{Conclusions}
Two-mode spin squeezed states are important states to produce for quantum
teleportation. Here we have shown that it is possible to produce states very
close to the optimal TMSS states by adapting the feedback for single-mode spin
squeezing considered by TMW. These states are not conditioned on the measurement
record, in contrast to the conditional two-mode spin squeezing discussed in
Ref.\ \cite{Julsgaard}.

Using the simple feedback scheme (\ref{feed}) it is possible to obtain states
that are quite close to optimal TMSS states for small times, but strongly
diverge from these states at later times. In particular, for spins above about
5 it is possible to obtain states very close to the TMSS states considered for
teleportation in Ref.\ \cite{Berry}. An analytic feedback scheme (\ref{analyt})
also gives similar results. This feedback scheme is more practical
experimentally, as the appropriate feedback strength to be used is easily
calculated.

We have derived the optimal feedback that produces the maximum possible
spin squeezing. This feedback is also applicable to the case of feedback for
single-mode spin squeezing. This feedback gives states even closer to the
optimal TMSS states.

We have also derived a Hamiltonian for the two-mode case that is equivalent to
the two-axis countertwisting Hamiltonian introduced in Ref.\ \cite{Kit}. This
Hamiltonian produces spin squeezing, but not as strongly as the measurements
with feedback.

In the case of spin 1/2 both the feedback (except for the analytic feedback)
and the countertwisting Hamiltonian produce optimal TMSS states. These states
are equivalent to optimal single-mode spin squeezed states if the two modes are
considered as a single spin-1 system. In the single-mode case optimal spin
squeezed states are produced both by feedback and by the countertwisting
Hamiltonian.

\acknowledgments
The authors acknowledge valuable discussions with Laura Thomsen and Howard
Wiseman. We are also grateful for constructive criticism from Eugene Polzik.
This project has been supported by an Australian Research Council Large Grant.

\appendix
\section{Relation of variables to experiment}
\label{sec:apa}
Here we give further explanation of the variables introduced in
Eq.\ (\ref{Hint}). The coupling constant $a$ is given by
\begin{equation}
a=\frac{\sigma}{A(I+1/2)}\frac{\gamma}{\Delta} \alpha_v,
\end{equation}
where $\sigma$ is the resonant absorption cross section, $A$ is the area of the
transverse cross section of the light beam, $I$ is the nuclear spin, $\gamma$ is
the spontaneous emission rate of the upper atomic level, $\Delta$ is the
detuning, and $\alpha_v$ is the dynamic vector polarizability. We have omitted
the bounds on the integral (\ref{Hint}) for greater generality, so we may apply
this expression to multiple samples. Explicit bounds are unnecessary, as there
is no contribution to the integral from regions where there are no atoms.

The continuous spin operators for the sample are defined as
\begin{equation}
J_{k}(z,t) = \lim_{\delta z \to 0} \frac 1{\delta z} \sum_{\mu} \tfrac
12 \sigma_{k}^{\mu},
\end{equation}
where $k \in \{x,y,z\}$ and $\sigma_{k}^{\mu}$ is the Pauli operator
for particle $\mu$. The sum is over all particles between $z$ and $z+\delta z$
over the cross section of the sample. The operator for the entire sample is
obtained by integrating over $z$.

The field is described by the continuous-mode annihilation operators
$a_{k}(z,t)$, where $k=x$ and $y$ for the $x$ polarized and $y$ polarized modes,
respectively. The instantaneous Stokes parameters are
\begin{align}
S_x(z,t) & =\tfrac 12[a_x\dg(z,t)a_x(z,t)-a_y\dg(z,t)a_y(z,t)], \nn \\
S_y(z,t) & =\tfrac 12[a_x\dg(z,t)a_y(z,t)+a_y\dg(z,t)a_x(z,t)], \nn \\
S_z(z,t) & =-\frac i2[a_x\dg(z,t)a_y(z,t)-a_y\dg(z,t)a_x(z,t)].
\end{align}
The Stokes vector for the entire pulse at position $z$ is given by
${\bf S}(z) = \int {\bf S}(z,t) dt$.

\section{Feedback for total spin 1}
\label{derivs}
Here we show that the feedback of either Eqs.\ (\ref{singlefeed}) or
(\ref{feed}) gives optimal spin squeezed states for a total spin of 1. To see
this, note first that the first term in Eq.\ (\ref{master}) is just the same as that
produced by the countertwisting Hamiltonian, and so will produce optimal states.
In order to show that the feedback produces optimal states, it remains to be
shown that the second term, ${\cal D}[c-iF]\rho$, does not alter the evolution.
Specifically, in the single-mode case
\begin{equation}
{\rm Tr}\left\{ J_x {\cal D}[J_z-i\Lambda J_y] \rho \right\} = -\tfrac 12
(1+\Lambda^2)\st{J_x} + \Lambda \st{J_z^2+J_y^2}.
\end{equation}
If the state $\rho$ is an optimal state, then $\st{J_z^2+J_y^2}=1$, so this
simplifies to
\begin{equation}
\label{zero1}
{\rm Tr}\left\{ J_x {\cal D}[J_z-i\Lambda J_y] \rho \right\} = -\tfrac 12
(1+\Lambda^2)\st{J_x} + \Lambda.
\end{equation}
Similarly we can show that
\begin{equation}
{\rm Tr}\left\{ J_z^2 {\cal D}[J_z-i\Lambda J_y] \rho \right\} = -\frac
{\Lambda}2 \st{J_x} + \Lambda^2 \st{J_x^2-J_z^2}.
\end{equation}
For optimal states, $\st{J_x^2}=1$, so this becomes
\begin{equation}
\label{zero2}
{\rm Tr}\left\{ J_z^2 {\cal D}[J_z-i\Lambda J_y] \rho \right\} = -\frac
{\Lambda}2 \st{J_x} + \Lambda^2 (1-\st{J_z^2}).
\end{equation}
It is simple to show that both Eqs.\ (\ref{zero1}) and (\ref{zero2}) will be
zero if Eq.\ (\ref{optrel}) is satisfied, and the feedback is given by
Eq.\ (\ref{singlefeed}).

Therefore, if the state is already in an optimal state, and the feedback as
given by Eq.\ (\ref{singlefeed}) is used, then the state will continue to be in
an optimal state. On the other hand, if some other feedback is used, then
optimal states will not be obtained. For example, the analytic feedback
considered by LMW does not give optimal states. In order to determine analytic
feedback that will give optimal states, note that the differential equations for
$\st{J_x}$ and $\st{J_z^2}$ are
\begin{align}
\frac{d}{dv} \st{J_x} & = \Lambda (2\st{J_z^2}-1), \nn \\
\frac{d}{dv} \st{J_z^2} & = -\frac{\Lambda}2 \st{J_x}.
\end{align}
Solving these using the feedback (\ref{singlefeed}) and Eq.\ (\ref{optrel}) gives
\begin{align}
\ip{J_x} & = \sqrt{2e^{-v}-e^{-2v}}, \nn \\
\ip{J_z^2} & = \tfrac 12 e^{-v}.
\end{align}
This implies that the value of $\Lambda$ should change with time as
\begin{equation}
\label{apeq}
\Lambda(v) = \frac 1{\sqrt{2e^{v}-1}}.
\end{equation}
This is quite different from the analytic expression applied for larger spins.

The case for feedback for two-mode spin squeezing is analogous to this, except
that $J_x$, $J_y$, and $J_z$ are replaced with $J_x^{(+)}$, $J_y^{(\Omega)}$,
and $J_z^{(\Omega)}$. This therefore shows that optimal states are obtained in
the one- and two-mode cases using the feedback given by Eqs.\ (\ref{singlefeed})
and (\ref{feed}), respectively.

\end{document}